\documentclass[twocolumn,twoside,slac_two]{revtex4}
\usepackage{graphicx}
\usepackage{fancyhdr}
\usepackage{amsmath}

\def\frac#1#2{{\textstyle{{#1}\over {#2}}}}

\def\Re{\hbox{Re}\,}
\def\Im{\hbox{Im}\,}

\def\ni{\noindent}
\newcommand{\beq}{\begin{equation}}
\newcommand{\eeq}{\end{equation}}
\newcommand{\bea}{\begin{eqnarray}}
\newcommand{\eea}{\end{eqnarray}}
\newcommand{\bit}{\begin{itemize}}
\newcommand{\eit}{\end{itemize}}

\newcommand{\nn}{\nonumber}
\newcommand{\B}{\Big}
\newcommand{\BB}{\big}

\newcommand{\W}{\widetilde}

\newcommand{\iM}{\begin{pmatrix}}
\newcommand{\fM}{\end{pmatrix}}

\def\HL{\delta h}

\def\H0#1{\frac{(m_{\text{SM}}^2)_{#1}}{2E}}
\def\H#1#2{\mathcal{H}^{(#1)}_{#2}} 
\newcommand{\isq}{\begin{subequations}\begin{align}}
\newcommand{\fsq}{\end{align}\end{subequations}}

\begin{document}

\title{Long-baseline neutrino experiments as tests for Lorentz violation}

%

\author{Jorge S. D\'iaz}
\affiliation{Physics Department, Indiana University, Bloomington, IN 47405, USA}

\begin{abstract}
Precise tests of Lorentz invariance can be executed using neutrino oscillations, which can provide sensitive measurements of suppressed signals of new physics. This talk describes the neutrino sector of the Standard-Model Extension, which represents a general modification of the Standard Model of massive neutrinos to include Lorentz and CPT violation. Attainable sensitivities as well as a framework to search for these violations are presented for existing and future long-baseline neutrino experiments. The applicability of this framework to short-baseline experiments is also discussed.
\end{abstract}

\maketitle

\thispagestyle{fancy}


\section{Introduction}
Present and future neutrino-oscillation experiments have been designed to measure unknown parameters describing standard neutrino behavior. Nevertheless, results from these experiments could also be used as precise tests for Lorentz symmetry, which is one of the cornerstones of our two more successful descriptions of nature: General Relativity and the Standard Model (SM). 

The lack to date of compelling evidence of any breaking of Lorentz invariance has left unaffected the interest on possible violations of this fundamental symmetry. On the contrary, enormous efforts have been made on both theoretical and experimental fronts developing models to describe consequences of any deviation from exact symmetry and searching for its direct effects in the laboratory.

Our limited current understanding of natural phenomena can be the consequence of not only the fact that we do not have yet a complete and unified theory able to describe gravity and quantum effects as a whole but also the limited energies capable of being produced in our experiments. Independent of the underlying theory, possible mechanisms that could lead to Lorentz and CPT violations at the Planck scale ($M_P\simeq10^{19}$ GeV) \cite{KosteleckySamuel1989} motivated the development of a theoretical description of these violations and their effects on low-energy regimes that can be studied in current experiments. The Standard-Model Extension (SME) is an effective field theory that, incorporating the SM and General Relativity, describes all possible Lorentz and CPT violations \cite{SME}. 
This framework allows for direct searches of departures from exact symmetry in different sectors of the SM. Lorentz violation is implemented as observer scalars constructed by contracting Lorentz-violating operators with controlling coefficients. Phenomenological effects measurable with current energy capabilities would be produced by these coefficients for Lorentz and CPT violation. Current experimental searches have led to sensitive constraints of these coefficients in matter, gauge, and gravitational sectors \cite{Tables}.

In this talk, we review the construction of a theory describing the experimental implications of departures from Lorentz and CPT invariance for neutrino oscillations developed in collaboration with Alan Kosteleck\'y and Matthew Mewes \cite{DiazKosteleckyMewes2009}. The interferometric nature of these oscillations make them highly sensitive to Planck-scale suppressed coefficients for Lorentz violation in the neutrino sector of the SME. In the present work, these coefficients are spacetime constants which can be interpreted as vacuum expectation values gotten by background tensor fields in the underlying theory, spontaneously breaking Lorentz symmetry. The disregarded Nambu-Goldstone modes could play fundamental roles when gravity is included, such as the graviton, the photon in Einstein-Maxwell theory, or spin-independent forces \cite{NGmodes}. However, these possibilities lie outside of the scope of the present work \cite{GravPapers}.

\section{Lorentz and CPT violation in neutrinos}

\label{Theory}

Oscillations of three left-handed neutrinos are described by the effective hamiltonian 

\beq\label{h_eff}
(h_{\text{eff}})_{ab}=(h_0)_{ab}+\HL_{ab},
\eeq

\ni
where $h_0$ is the standard three-neutrino massive hamiltonian \cite{PDG2008}, whereas $\HL$ includes Lorentz- and CPT-violating contributions. The indices span the three neutrino flavors $a,b,\ldots=e,\mu,\tau$. The Lorentz-violating term $\HL$ is given by \cite{KosteleckyMewes2004a}

\beq\label{h_LV}
\HL_{ab} =\frac{1}{E}\BB[(a_L)^\alpha p_\alpha-(c_L)^{\alpha\beta} p_\alpha p_\beta\BB]_{ab},
\eeq

\ni
where $(a_L)^\alpha_{ab}$ and $(c_L)^{\alpha\beta}_{ab}$ are coefficients for Lorentz violation. The formers have mass dimensions whereas the latter are dimensionless. Oscillations acquire unconventional energy dependence given the dimensionalities of these coefficients. In the standard massive model $h_0$, oscillations are controlled by the dimensionless combination $\Delta m^2L/E$; the Lorentz-violating contribution (\ref{h_LV}) introduces the new combinations $(a_L)^\alpha_{ab} L$ and $(c_L)^{\alpha\beta}_{ab} LE$. Moreover, CPT-invariant effects are controlled by $(c_L)^{\alpha\beta}_{ab}$, whereas $(a_L)^\alpha_{ab}$ govern the CPT-violating ones. These last coefficients vanish when CPT symmetry holds. 

The oscillation of three right-handed antineutrinos is equivalent and all results can be applied using $(a_R)^\alpha_{\bar a\bar b}=-(a_L)^{\alpha\,*}_{ab}$ and $(c_R)^{\alpha\beta}_{\bar a\bar b}=(c_L)^{\alpha\beta\,*}_{ab}$ in Eq. (\ref{h_LV}), where barred indices span over the three antineutrino flavors $\bar a,\bar b,\ldots=\bar e,\bar\mu,\bar \tau$. The CPT theorem \cite{Greenberg} implies that the Lorentz-invariant hamiltonian for antineutrinos can be obtained using $(h_0)_{\bar a\bar b}=(h_0)_{ab}^*$. More details about antineutrino oscillations as well as possible neutrino-antineutrino mixing are widely discussed in Ref. \cite{DiazKosteleckyMewes2009}.

\label{SiderealVariations}

The presence of the neutrino four-momentum $p_\alpha\approx E(1;-\hat p)$ in the hamiltonian (\ref{h_LV}) explicitly shows that the direction of propagation plays an important role and hence it must be considered in any analysis involving Lorentz violation. The only exception to this is the case of isotropic models, in which the only nonzero coefficients are $(a_L)^T_{ab}$ and/or $(c_L)^{TT}_{ab}$. For Earth-based experiments, both source and detector rotate at angular frequency $\omega_\oplus\simeq2\pi/($23 h 56 min$)$, causing the neutrino direction to change with respect to the constant coefficients for Lorentz violation. This time dependence of the effective hamiltonian can be explicitly displayed as

\bea\label{sidereal_decomposition}
\HL_{ab}&=& (\mathcal{C})_{ab}+(\mathcal{A}_s)_{ab}\sin{\omega_\oplus T_\oplus}+(\mathcal{A}_c)_{ab}\cos{\omega_\oplus T_\oplus} \nn\\
&&+(\mathcal{B}_s)_{ab}\sin{2\omega_\oplus T_\oplus}+(\mathcal{B}_c)_{ab}\cos{2\omega_\oplus T_\oplus},
\eea 

\ni
where each amplitude depends on the coefficients $(a_L)^\alpha_{ab}$ and $(c_L)^{\alpha\beta}_{ab}$, the energy $E$, and the direction of the beam with respect to the Sun-centered celestial-equatorial frame \cite{KosteleckyMewesEM2002}. Although this sidereal decomposition shows that any analysis is highly experiment-dependent, the coefficients for Lorentz violation in any inertial frame can be related to the ones in the Sun-centered frame by an observer Lorentz transformation.

\section{Theory Classification}

\subsection{Negligible-Mass Theory}
\label{Negligible_Mass_Theory}

Observation of solar and atmospheric neutrinos can be understood using two mass-squared differences; however, the signal detected by the Liquid Scintillator Neutrino Detector (LSND) continues as a puzzling result that cannot be accommodated within this picture \cite{LSND2001}. The bicycle model \cite{KosteleckyMewes2004b} was the first attempt to include the LSND signal as a Lorentz-violating effect only, using the minimal SME. Having only two parameters and no neutrino masses, this model is consistent with atmospheric data from the Super-Kamiokande experiment \cite{Messier2005}, whose oscillations emerge from a direction-dependent pseudomass. Nonetheless, this model and its generalization were excluded by combining data from solar, long-baseline, and reactor experiments \cite{BargerMarfatiaWhisnant2007}.

Even though neutrino masses seem to play an important role describing oscillations, mass effects are not significant when $\Delta m^2L\ll E$. As a result, under suitable experimental conditions, mass effects can be irrelevant compared to the ones produced by $\HL$ in the oscillation probability \cite{KosteleckyMewes2004c}. In such cases, we can neglect $h_0$ in Eq. (\ref{h_eff}) and write the effective hamiltonian as $h_\text{eff}\approx\HL$. At leading order, the oscillation probability becomes

\beq
P_{\nu_b\rightarrow\nu_a}\simeq L^2|\HL_{ab}|^2,\quad a\neq b,
\eeq

\ni
where a short baseline $L$ compared to the oscillation length $1/h_\text{eff}$ is required for the validity of this expression. Most of the short-baseline experiments fall in this category; however, a relatively long-baseline experiment (hundreds of kilometers) using a high-energy beam (hundreds of GeV) would still be well described by this negligible-mass theory. It is important to notice that the applicability of any theory for the analysis of a given experiment depends on the location of the experiment in $L$-$E$ space \cite{DiazKosteleckyMewes2009}.

Anisotropic coefficients for Lorentz violation will lead to sidereal variations of the oscillation probability controlled by (\ref{sidereal_decomposition}). The structure of this time-dependent oscillation probability can be used to extract the coefficients for Lorentz violation contained in the sidereal amplitudes.

To date, two sets of coefficients have been constrained using this method by LSND \cite{LSND2005} and the Main Injector Neutrino Oscillation Search (MINOS) using its near detector (ND) \cite{MINOSND2008}. Studying different oscillation channels, LSND and MINOS ND constrained the real parts of the same coefficients for different neutrino flavors: \Re$(a_L)^\alpha_{\bar e\bar\mu}<10^{-19}$ GeV, $\Re(c_L)^{\alpha\beta}_{\bar e\bar\mu}<10^{-17}$ and $\Re(a_L)^\alpha_{\mu\tau}<10^{-20}$ GeV, $\Re(c_L)^{\alpha\beta}_{\mu\tau}<10^{-21}$, respectively. The longer baseline and higher energy of MINOS ND helped to increase its sensitivity compared to the LSND constraints. Results from these experimental analyses are consistent with Planck-scale suppression expected from quantum-gravity effects.

\subsection{Hybrid Models}

Models with mass terms for a subset of neutrinos are called hybrid. As an extension of the bicycle model, the tandem model \cite{KatoriKosteleckyTayloe2006} is the only example of these kind of models to date. It includes one CPT-even and one CPT-odd coefficient for Lorentz violation and one neutrino mass. 
This rotationally invariant model based on the minimal SME can accommodate the LSND anomaly and is compatible with features described by the three-neutrino massive model. The values of the two coefficients for Lorentz violation are chosen to be real and consistent with Planck-scale suppression and the LSND analysis \cite{LSND2005}. The mass parameter respects the cosmological constraint and is compatible with a seesaw origin. Oscillation probabilities within this model are consistent with experimental data. In particular, the null signal for the Karlsruhe-Rutherford Medium Energy Neutrino (KARMEN) \cite{KARMEN2002} experiment is understood as a consequence of its short baseline. A very novel prediction of this model is a low-energy excess for the Mini Booster Neutrino Experiment (MiniBooNE) prior to its observation \cite{MiniBooNE}. In spite of the fact that the predicted signal is lower than the observed, this remarkable prediction shows that very realistic models can be built based on the SME and the interesting capabilities of these hybrid models.

\subsection{Perturbative Lorentz and CPT Violation}
\label{Perturbative_Theory}

Experiments lying in the $L$-$E$ space where neutrino masses become relevant will not be well described by the theory presented in \ref{Negligible_Mass_Theory}. In this section we present the other sector of the general theory presented in section \ref{Theory}, where the mass contributions are dominant in the oscillation probability over the Lorentz-violating ones and all neutrinos have conventional masses. For details and explicit derivation see Ref. \cite{DiazKosteleckyMewes2009}, on which this section is based.

Using perturbation theory, we can describe the effects of $\HL$ on the background solutions that diagonalize the Lorentz-invariant hamiltonian $h_0$, which includes matter effects if necessary. Writing the transition amplitude as a series expansion $S_{ab}=S^{(0)}_{ab}+S^{(1)}_{ab}+\cdots$, the first-order correction to the oscillation probability is given by

\bea\label{P0_P1}
P_{\nu_b\rightarrow\nu_a}^{(1)}&=&2L\,\Im\BB((S^{(0)}_{ab})^*\H{1}{ab}\BB).
\eea

\ni
The quantities $\H{1}{ab}$ are linear combinations of the Lorentz-violating hamiltonian $\HL_{cd}$ and a collection of experimental-dependent factors tabulated in Ref. \cite{DiazKosteleckyMewes2009}. For observations of a given oscillation channel $\nu_b\rightarrow\nu_a$, we expect corrections produced by $\HL_{ab}$; however, Eq. (\ref{P0_P1}) reveals that all the components of $\HL_{cd}$ in flavor space contribute. Linearity leads to construct quantities $(\W a_L)^\alpha_{ab}$ and $(\W c_L)^{\alpha\beta}_{ab}$ as combinations of the original coefficients for Lorentz violation $(a_L)^\alpha_{ab}$ and $(c_L)^{\alpha\beta}_{ab}$, respectively. The new coefficients have the same structure than $\H{1}{ab}$; hence, we can write the components of this last quantity in the form of Eq. (\ref{h_LV}), replacing $(a_L)^\alpha_{ab}$ and $(c_L)^{\alpha\beta}_{ab}$ by their intermediate combinations $(\W a_L)^\alpha_{ab}$ and $(\W c_L)^{\alpha\beta}_{ab}$. The sidereal decomposition (\ref{sidereal_decomposition}) can then be used, which allows a similar decomposition of the oscillation probability (\ref{P0_P1}) given by

\bea
&&\!\!\!\!\!\!\!\!
\frac{P_{\nu_b\rightarrow\nu_a}^{(1)}}{2L}\,=\,\Im\BB((S^{(0)}_{ab})^*\H{1}{ab}\BB) \nn\\
&&\!\!\!\!\!\!\!\!
\quad=(P_{\mathcal{C}})_{ab}^{(1)}
+(P_{\mathcal{A}_s})_{ab}^{(1)}\sin{\omega_\oplus T_\oplus}%
+(P_{\mathcal{A}_c})_{ab}^{(1)}\cos{\omega_\oplus T_\oplus} \nn\\
&&\quad\quad
+(P_{\mathcal{B}_s})_{ab}^{(1)}\sin{2\omega_\oplus T_\oplus}
+(P_{\mathcal{B}_c})_{ab}^{(1)}\cos{2\omega_\oplus T_\oplus},
\eea

\ni
where the sidereal amplitudes are explicitly given in Ref. \cite{DiazKosteleckyMewes2009}. These amplitudes are expected to be tiny since they are linear combinations of the coefficients for Lorentz violation; nonetheless, we can see that their minuscule effects can be enhanced by large baselines. Section \ref{Negligible_Mass_Theory} discussed the fact that the applicability of a theory depends on the location of a given experiment in the $L$-$E$ space. Most of the long-baseline experiments lie in the appropriate region where the perturbative theory is applicable; however, short-baseline experiments working with low-energy neutrinos are also well described by this theory. Another important factor that affects the applicability of the theory to a given experiment is the value of $\theta_{13}$. If this mixing angle is small or zero, the mass mixing involves only two generations and a two-flavor limit of the perturbative theory can be applied \cite{DiazKosteleckyMewes2009}.

\section{Long-baseline Neutrino Experiments}

Most of the future long-baseline neutrino experiments aim to search for precise measurements of $\theta_{13}$ and possible signals of CP violation in the neutrino sector. Nevertheless, more physics can be extracted using these as well as present experiments and their data. The theory presented above can be used to search for possible signals of Lorentz and CPT violation and in this section we explore this possibility.

We apply the theory to eight long-baseline experiments namely: KEK to Super-Kamiokande experiment (K2K, $L\simeq250$ km) \cite{K2K}, MINOS far detector (FD, $L\simeq730$ km) \cite{MINOS}, the Oscillation Project with Emulsion-Tracking Apparatus (OPERA, $L\simeq730$ km) \cite{OPERA}, the Imaging Cosmic and Rare Underground Signal experiment (ICARUS, $L\simeq730$ km) \cite{ICARUS}, the NuMI Off-Axis $\nu_e$ Appearance experiment (NO$\nu$A, $L\simeq810$ km) \cite{NOvA}, the Tokai to Kamioka experiment (T2K, $L\simeq300$ km) \cite{T2K}, the Deep Underground Science and Engineering Lab experiment (DUSEL, $L\simeq1300$ km) \cite{DUSEL}, and the Tokai to Kamioka and Korea experiment (T2KK, $L\simeq1000$ km) \cite{T2KK}.

\subsection{$\nu_e$ appearance}

Let us first consider a nonzero value for $\theta_{13}$; thus, according to the discussion at the end of last section, a three-flavor analysis is required for our eight experiments.

The three-neutrino massive model \cite{PDG2008} describes neutrino oscillations using six parameters: two mass-squared differences, three mixing angles, and one CP-violating phase. We assume values for these parameters that are consistent with current data \cite{PDG2008}. Even though we use $\delta\simeq0^\circ$ in the mixing matrix, CP violation could still arise from Lorentz- and CPT-violating terms in (\ref{h_LV}).

The correction to the $\nu_e$-appearance oscillation probability in a $\nu_\mu$ beam depends on the combinations $(\W a_L)^\alpha_{e\mu}$ and $(\W c_L)^{\alpha\beta}_{e\mu}$ contained in the sidereal amplitudes. Since these amplitudes are different for each experiment, the correction to the oscillation probability will be different as well; consequently, the measurement of a particular combination of coefficients for Lorentz violation can be done by each experiment. This feature implies that the coverage of the whole coefficient space depends on multiple experiments. For details on all the sidereal amplitudes for the eight experiments mentioned above see Ref. \cite{DiazKosteleckyMewes2009}.

\subsection{$\nu_\mu$ disappearance}

The use of $\nu_\mu$ beams allows the search for their disappearance. At the high energies of our eight experiments, the mixing is mainly between $\nu_\mu$ and $\nu_\tau$ and oscillations are well approximated by a two-flavor system. This limit can also be considered if $\theta_{13}$ is small or zero. In these cases the disappearance oscillation probability reduces to $P_{\nu_\mu\rightarrow\nu_X}=P_{\nu_\mu\rightarrow\nu_\tau}$, in other words, in a two-flavor system $\nu_\mu$ disappearance is equivalent to $\nu_\tau$ appearance. In this approximation, the Lorentz-invariant hamiltonian $h_0$ involves only one relevant mass-squared difference and the mixing matrix is a simple rotation by angle $\theta_{23}$. Using maximal mixing on this two-flavor system, the correction to the $\nu_\mu$ disappearance oscillation probability takes the simple form

\beq
P^{(1)}_{\nu_\mu\rightarrow\nu_\tau}\approx\Re (\HL_{\mu\tau})L\,\sin(\Delta m^2_\text{atm}L/2E),
\eeq

\ni
where only the real part of one of the components of $\HL$ contributes. This last expression can also be obtained by taking the limit $\Delta m^2_\odot\ll\Delta m^2_\text{atm}$ and neglecting mixing other than $\theta_{23}$ in the three-flavor analysis presented before. The sidereal decomposition can now be applied and different combinations of coefficients for Lorentz violation for each experiments can be found. The main differences compared to the three-flavor case are the dependence on the real part of $\HL_{\mu\tau}$ only and the direct presence of the coefficients for Lorentz violation $(a_L)^\alpha_{\mu\tau}$ and $(c_L)^{\alpha\beta}_{\mu\tau}$ instead of the combinations $(\W a_L)^\alpha_{\mu\tau}$ and $(\W c_L)^{\alpha\beta}_{\mu\tau}$. This implies that any search for sidereal amplitudes leads to a direct measurement of the effects of the coefficients in Eq. (\ref{h_LV}). For details on the sidereal amplitudes for each experiment see Ref. \cite{DiazKosteleckyMewes2009}.

\subsection{Sensitivities}

The study of the two oscillation channels presented here is part of the programs both current and proposed by the different long-baseline experiments used in our analyses. Estimated sensitivities to the sidereal amplitudes are of order $10\%/2L$, which can be obtained supposing that each experiment can detect a 10\% sidereal variation in the corresponding oscillation probability. Simple calculation reveals that sidereal amplitudes would be of order $10^{-23}$ GeV. This value turns out to be consistent with Planck-scale suppression effects expected for Lorentz and CPT violation. Moreover, it indicates that the mentioned experiments using this theory could improve by a couple orders of magnitude the current values of coefficients for Lorentz violation. Additionally, the analysis involving three flavors allows access to most of the coefficients via the intermediate combinations $(\W a_L)^\alpha_{ab}$ and $(\W c_L)^{\alpha\beta}_{ab}$. On the other hand, two-flavor analyses restrict the possible coefficients to be constrained to real parts and flavor indices $\{ab\}=\{\mu\tau\}$ only. Nevertheless, disappearance searches have better statistics.

Coefficients for Lorentz violation having spacetime indices $T$, $Z$, $TT$, $TZ$, and $ZZ$ are challenging to be measured because they are contained in the amplitudes $(P^{(1)}_\mathcal{C})_{ab}$, which do not present variations with time. Time dependence of the other four terms in the sidereal decomposition constitutes a signal to be sought in experimental data. In the case of absence of any sidereal variation in the data, constraints can be found on the coefficients for Lorentz violation, in the same way that SME coefficients are constrained in other sectors \cite{OtherSectors}.

It is important to emphasize again that despite the fact that we are using long-baseline experiments, the perturbative theory presented in section \ref{Perturbative_Theory} can also be used to analyze data from short-baseline experiments, as long as their location in the $L$-$E$ space is appropriate to the perturbation to be valid. For instance, reactor experiments searching for $\bar\nu_e$ disappearance are located in a valid region in the $L$-$E$ space where neutrino masses become relevant and the perturbative theory is applicable.

\section{CPT violation}

To date, there are no evidence of neither Lorentz nor CPT violations in nature. Theoretically, these two symmetries are deeply related. In a framework consistent with quantum field theory, any CPT violation implies the also breaking of Lorentz symmetry \cite{Greenberg}. On the experimental front, some of the present and future experiments have the potential to search directly for CPT violation in the neutrino sector. Experiments able to change the polarity of their horns and focus positive and negative charged mesons into the decay pipe can run in both neutrino and antineutrinos modes. This feature allows the direct study of neutrino oscillations and their CP conjugates. The connection between CP and CPT depends on the T-transformation properties of the system. The oscillation probability is T invariant when the mixing matrix is real. This condition is satisfied when the CP phase $\delta$ vanishes. Although a zero CP phase in the mixing matrix leads to CP invariant physics in the standard neutrino model \cite{PDG2008}, CP violation could arise from CPT-odd operators violating Lorentz symmetry in (\ref{h_LV}). In a three-flavor description, a zero $\theta_{13}$ also leads to real mixing. In a two-flavor one, the mixing is always real and the system is T invariant.

Asymmetries are usually defined to measure departures from a given symmetry. In our case, a CPT asymmetry can be defined as

\vspace{-0.2cm}

\beq\label{A_CPT}
\mathcal{A}^{CPT}_{ab}=\frac{P_{\nu_b\rightarrow\nu_a}-P_{\bar\nu_a\rightarrow\bar\nu_b}}{P_{\nu_b\rightarrow\nu_a}+ P_{\bar\nu_a\rightarrow\bar\nu_b}}.
\eeq

\ni
A CP asymmetry $\mathcal{A}^{CP}_{ab}$ can be defined in a similar form. For T-invariant systems $\mathcal{A}^{CP}_{ab}=\mathcal{A}^{CPT}_{ab}$. Restricting our attention to two-flavor systems, we can write the asymmetry (\ref{A_CPT}) in terms of coefficients for Lorentz violation. Depending on the region of parameter space where the experiment operates, two asymmetries can be defined: $\mathcal{A}^{CPT}_{\mu\mu}$ and $\mathcal{A}^{CPT}_{\mu\tau}$. For instance, at first order the first of these asymmetries is given by

\vspace{-0.2cm}

\beq\label{A_CPT(h)}
\mathcal{A}^{CPT}_{\mu\tau}\approx2L\,\cot\B(\frac{\Delta m^2_\text{atm}L}{4E}\B)\,\Re (\HL)_{\mu\tau}^{CPT} ,
\eeq

\ni
where $(\HL)_{\mu\tau}^{CPT}$ is the CPT-odd part of $\HL_{\mu\tau}$. Using Eq. (\ref{sidereal_decomposition}), we can similarly perform a sidereal decomposition of this asymmetry. Equation (\ref{A_CPT(h)}) has the same structure than the corresponding CPT-violation parameter used in neutral meson systems \cite{KosteleckyMesons}. An interesting aspect is that these asymmetries offer the possibility to constrain the real parts of $(a_L)^T_{\mu\tau}$ and $(a_L)^Z_{\mu\tau}$ by taking an average in time, which would otherwise be challenging given the time independence of $(P^{(1)}_\mathcal{C})_{ab}$. 

In order to be consistent with quantum field theory, in this study of CPT violation using neutrinos and antineutrinos we have avoided the commonly used phenomenological approach of comparing masses of particles and their antiparticles. The breaking of Lorentz symmetry accompanying any CPT violation \cite{Greenberg} implies that parameters measuring departures from the exact symmetry cannot be Lorentz scalars; on the contrary, they must depend on particle energy and momentum. Particle momentum introduces direction dependence; consequently, the sidereal decomposition of the hamiltonian becomes a powerful experimental tool to study any asymmetry. Additionally, we have seen that any analysis is highly experiment dependent. Different experiments will have uncorrelated sensitivities to coefficients for Lorentz and CPT violation because the propagation of neutrino beams in different directions lead to unlike couplings to the background fields; therefore, analysis involving results from different experiments must be treated separately.

\section{Summary}

In this talk, we review the basic theory describing Lorentz and CPT violation in neutrino oscillations within the SME. Non-stardard energy dependence as well as sidereal variations on the oscillation probabilities arise as potential signals to be measured in present and future experiments. Two classes of theories have been classified according to their neutrino-mass content, which depends on the baseline and energy range covered by a given experiment. For experiments in which neutrino masses are irrelevant for the oscillation probability, the negligible mass theory discussed in section \ref{Negligible_Mass_Theory} properly describes oscillations produced by coefficients for Lorentz violation. For experiments located in a region of the $L$-$E$ space where neutrino masses dominate the oscillation probability, the perturbative theory presented in section \ref{Perturbative_Theory} describes the possible corrections introduced by Lorentz and CPT violations to the standard massive neutrino model. This perturbative theory is applied to eight experiments laying in the appropriate region of the $L$-$E$ space: K2K, MINOS FD, OPERA, ICARUS, NO$\nu$A, T2K, DUSEL, and T2KK. Half of the experiments in this list are either currently taking data or done with their data acquisition. The other half are experiments under construction and proposed projects. In any case, all of them offer excellent sensitivities to coefficients for Lorentz and CPT violation, improving by several orders of magnitude the current constrains. The long baselines enhance the effects of these coefficients, which are expected to be suppressed by a factor $M_P$.

A simple model built upon two coefficients for Lorentz violation and one neutrino mass is also reviewed in this talk. The tandem model corresponds to a realistic example of hybrid models that is consistent with solar, atmospheric, accelerator, reactor, and LSND data. Remarkably, it predicted a low-energy excess in the MiniBooNE experiment before to its observation, presenting the potential capabilities of hybrid models.

Finally, a framework for CPT violation is presented for experiments able to run in both neutrino and antineutrino modes. The CPT asymmetries explicitly present the energy and momentum dependence expected for parameters measuring CPT violation, which leads to the corresponding sidereal decomposition. Furthermore, the asymmetry time-average allows access to some coefficients that are otherwise challenging to be constrained. In brief, present and future experiments offer excellent sensitivities on searches for Lorentz and CPT violation and most of the coefficient space can be covered using the procedures presented in this talk.



\begin{thebibliography}{9}   

\bibitem{KosteleckySamuel1989}   
  V.A.\ Kosteleck\'y and S.\ Samuel,
  Phys.\ Rev.\  D {\bf 39}, 683 (1989);
  V.A.\ Kosteleck\'y and R.\ Potting,
  Nucl.\ Phys.\  B {\bf 359}, 545 (1991).



\bibitem{SME}
  D.\ Colladay and V.A.\ Kosteleck\'y,
  Phys.\ Rev.\  D {\bf 55}, 6760 (1997);
  Phys.\ Rev.\  D {\bf 58}, 116002 (1998);
  V.A.\ Kosteleck\'y,
  Phys.\ Rev.\  D {\bf 69}, 105009 (2004).



\bibitem{Tables}
  {\it Data Tables for Lorentz and CPT Violation,}
  V.A.\ Kosteleck\'y and N.\ Russell, 2009 edition,
  arXiv:0801.0287v2.

  


\bibitem{DiazKosteleckyMewes2009}   
  J.S.\ D\'iaz, V.A.\ Kosteleck\'y and M.\ Mewes,
  Phys.\ Rev.\  D, in press,
  arXiv:0908.1401.
  
  
\bibitem{NGmodes}
  V.A.\ Kosteleck\'y and R.\ Potting,
  Gen.\ Rel.\ Grav.\  {\bf 37}, 1675 (2005);
  Phys.\ Rev.\  {\bf D79}, 065018 (2009);
  V.A.\ Kosteleck\'y and S.\ Samuel,
  Phys.\ Rev.\  D {\bf 40}, 1886 (1989);
  Phys.\ Rev.\ Lett.\  {\bf 63}, 224 (1989);
  R.\ Bluhm and V.A.\ Kosteleck\'y,
  Phys.\ Rev.\  D {\bf 71}, 065008 (2005);
  O.\ Bertolami and J.\ Paramos,
  Phys.\ Rev.\  D {\bf 72}, 044001 (2005);
  M.D.\ Seifert,
  Phys.\ Rev.\  D {\bf 79}, 124012 (2009);
  V.A.\ Kosteleck\'y and J.\ Tasson,
  Phys.\ Rev.\ Lett.\  {\bf 102}, 010402 (2009).

  
  \bibitem{GravPapers}
  For recent developments in the gravity and photon sectors of the SME see:
  Q.G.\ Bailey,
  Phys.\ Rev.\  D {\bf 80}, 044004 (2009);
  V.A.\ Kosteleck\'y, N.\ Russell and J.\ Tasson,
  Phys.\ Rev.\ Lett.\  {\bf 100}, 111102 (2008);
  R.\ Bluhm {\it et al.},
  Phys.\ Rev.\  D {\bf 77}, 065020 (2008);
  Q.G.\ Bailey and V.A.\ Kosteleck\'y,
  Phys.\ Rev.\  D {\bf 74}, 045001 (2006);
  %
  M.D.\ Seifert,
  arXiv:0909.3118;
  V.A.\ Kosteleck\'y and M.\ Mewes,
  Phys.\ Rev.\  D {\bf 80}, 015020 (2009);
  Astrophys.\ J.\  {\bf 689}, L1 (2008);
  B.\ Altschul,
  Phys.\ Rev.\ Lett.\  {\bf 98}, 041603 (2007);
  Q.G.\ Bailey and V.A.\ Kosteleck\'y,
  Phys.\ Rev.\  D {\bf 70}, 076006 (2004);
  R.\ Jackiw and V.A.\ Kosteleck\'y,
  Phys.\ Rev.\ Lett.\  {\bf 82}, 3572 (1999).


\bibitem{PDG2008}
  C.\ Amsler {\it et al.}  (Particle Data Group),
  Phys.\ Lett.\  B {\bf 667}, 1 (2008).
  

\bibitem{KosteleckyMewes2004a}
  V.A.\ Kosteleck\'y and M.\ Mewes,
  Phys.\ Rev.\  D {\bf 69}, 016005 (2004).
  
\bibitem{Greenberg}
  O.W.\ Greenberg,
  Phys.\ Rev.\ Lett.\  {\bf 89}, 231602 (2002).

\bibitem{KosteleckyMewesEM2002}
  V.A.\ Kosteleck\'y and M.\ Mewes,
  Phys.\ Rev.\  D {\bf 66}, 056005 (2002).
  


\bibitem{LSND2001}
  LSND Collaboration, A.\ Aguilar {\it et al.},
  Phys.\ Rev.\  D {\bf 64}, 112007 (2001).
  

\bibitem{KosteleckyMewes2004b}
  V.A.\ Kosteleck\'y and M.\ Mewes,
  Phys.\ Rev.\  D {\bf 70}, 031902 (2004).
  
  
\bibitem{Messier2005}
  M.D.\ Messier, 
  in V.A.\ Kosteleck\'y, ed., 
  {\it CPT and Lorentz Symmetry III}, World Scientific, Singapore, 2005.



\bibitem{BargerMarfatiaWhisnant2007}
  V.\ Barger, D.\ Marfatia and K.\ Whisnant,
  Phys.\ Lett.\  B {\bf 653}, 267 (2007).


\bibitem{KosteleckyMewes2004c}
  V.A.\ Kosteleck\'y and M.\ Mewes,
  Phys.\ Rev.\  D {\bf 70}, 076002 (2004).



\bibitem{LSND2005}
  LSND Collaboration, L.B.\ Auerbach {\it et al.},
  Phys.\ Rev.\  D {\bf 72}, 076004 (2005).

\bibitem{MINOSND2008}
  MINOS Collaboration, P.\ Adamson {\it et al.},
  Phys.\ Rev.\ Lett.\  {\bf 101}, 151601 (2008).


\bibitem{KatoriKosteleckyTayloe2006}
  T.\ Katori, V.A.\ Kosteleck\'y and R.\ Tayloe,
  Phys.\ Rev.\  D {\bf 74}, 105009 (2006).


\bibitem{KARMEN2002}
  KARMEN Collaboration, B.\ Armbruster {\it et al.},
  Phys.\ Rev.\  D {\bf 65}, 112001 (2002).



\bibitem{MiniBooNE}
  MiniBooNE Collaboration, A.A.\ Aguilar-Arevalo {\it et al.},
  Phys.\ Rev.\ Lett.\  {\bf 98}, 231801 (2007),
%
  Phys.\ Rev.\ Lett.\  {\bf 102}, 101802 (2009).



\bibitem{K2K}
  K2K Collaboration, M.H.\ Ahn {\it et al.},
  Phys.\ Rev.\  D {\bf 74}, 072003 (2006).


\bibitem{MINOS}
  MINOS Collaboration, D.G.\ Michael {\it et al.},
  Nucl.\ Instrum.\ Meth.\  A {\bf 596}, 190 (2008).

\bibitem{OPERA}
  OPERA Collaboration, M.\ Guler {\it et al.},
  preprint CERN-SPSC-2000-028 (2000).  
  

\bibitem{ICARUS}
  ICARUS Collaboration, S.\ Amerio {\it et al.},
  Nucl.\ Instrum.\ Meth.\  A {\bf 527}, 329 (2004).


\bibitem{NOvA}
  NO$\nu$A Collaboration, D.S.\ Ayres {\it et al.},
   ``NO$\nu$A proposal to build a 30-kiloton off-axis detector to study neutrino,
  arXiv:hep-ex/0503053.


\bibitem{T2K}
  T2K Collaboration, Y.\ Itow {\it et al.},
  arXiv:hep-ex/0106019.


\bibitem{DUSEL}
  V.\ Barger {\it et al.},
  arXiv:0705.4396.


\bibitem{T2KK}
  K.\ Hagiwara, N.\ Okamura and K.i.\ Senda,
  Phys.\ Lett.\  B {\bf 637}, 266 (2006).



\bibitem{KosteleckyMesons}
  V.A.\ Kosteleck\'y,
  Phys.\ Rev.\  D {\bf 64}, 076001 (2001);
%
  Phys.\ Rev.\ Lett.\  {\bf 80}, 1818 (1998);
%
  Phys.\ Rev.\  D {\bf 61}, 016002 (2000).
%
  V.A.\ Kosteleck\'y and R.\ Potting,
  Phys.\ Rev.\  D {\bf 51}, 3923 (1995).

\bibitem{OtherSectors}
  B.\ Altschul,
  arXiv:0905.4346;
  Phys.\ Rev.\ Lett.\  {\bf 96}, 201101 (2006);
%
  M.A.\ Hohensee, R.\ Lehnert, D.F.\ Phillips and R.L.\ Walsworth,
  Phys.\ Rev.\ Lett.\  {\bf 102}, 170402 (2009);
  Phys.\ Rev.\  D {\bf 80}, 036010 (2009);
  R.\ Lehnert,
  Phys.\ Rev.\  D {\bf 68}, 085003 (2003);
  R.\ Lehnert and R.\ Potting,
  Phys.\ Rev.\ Lett.\  {\bf 93}, 110402 (2004).
  See also Ref. \cite{Tables}.
  




\end{thebibliography}

\end{document}